\documentclass[journal=ancac3]{achemso}
\usepackage[version=3]{mhchem}

\usepackage{epstopdf}
\usepackage[mathlines]{lineno}

\author{Pavel~A.~Dmitriev}
\affiliation[ITMO University]
{Laboratory of Nanophotonics and Metamaterials, ITMO University, St.~Petersburg 197101, Russia}

\author{Denis~G.~Baranov}
\affiliation[Moscow Institute of Physics and Technology]
{Moscow Institute of Physics and Technology, Dolgoprudny 141700, Russia}

\author{Valentin~A.~Milichko}
\affiliation[ITMO University]
{Laboratory of Nanophotonics and Metamaterials, ITMO University, St.~Petersburg 197101, Russia}

\author{Sergey~V.~Makarov}
\affiliation[ITMO University]
{Laboratory of Nanophotonics and Metamaterials, ITMO University, St.~Petersburg 197101, Russia}

\author{Ivan~S.~Mukhin}
\affiliation[ITMO University]
{Laboratory of Nanophotonics and Metamaterials, ITMO University, St.~Petersburg 197101, Russia}
\alsoaffiliation[St. Petersburg Academic University]
{Laboratory of Renewable Energy Sources, St. Petersburg Academic University, St.~Petersburg 194021, Russia}

\author{Anton~K.~Samusev}
\affiliation[ITMO University]
{Laboratory of Nanophotonics and Metamaterials, ITMO University, St.~Petersburg 197101, Russia}

\author{Alexander~E.~Krasnok}
\affiliation[ITMO University]
{Laboratory of Nanophotonics and Metamaterials, ITMO University, St.~Petersburg 197101, Russia}
\email{krasnokfiz@mail.ru}

\author{Pavel~A.~Belov}
\affiliation[ITMO University]
{Laboratory of Nanophotonics and Metamaterials, ITMO University, St.~Petersburg 197101, Russia}

\author{Yuri~S.~Kivshar}
\affiliation[ITMO University]
{Laboratory of Nanophotonics and Metamaterials, ITMO University, St.~Petersburg 197101, Russia}
\alsoaffiliation[Nonlinear Physics Centre]
{Nonlinear Physics Centre, Australian National University, Canberra ACT 2601, Australia}

\title[An \textsf{achemso} demo] {Resonant Raman Scattering from Silicon Nanoparticles Enhanced by Magnetic Response}


\begin{document}

\begin{tocentry}
\centering\includegraphics[width=0.99\columnwidth]{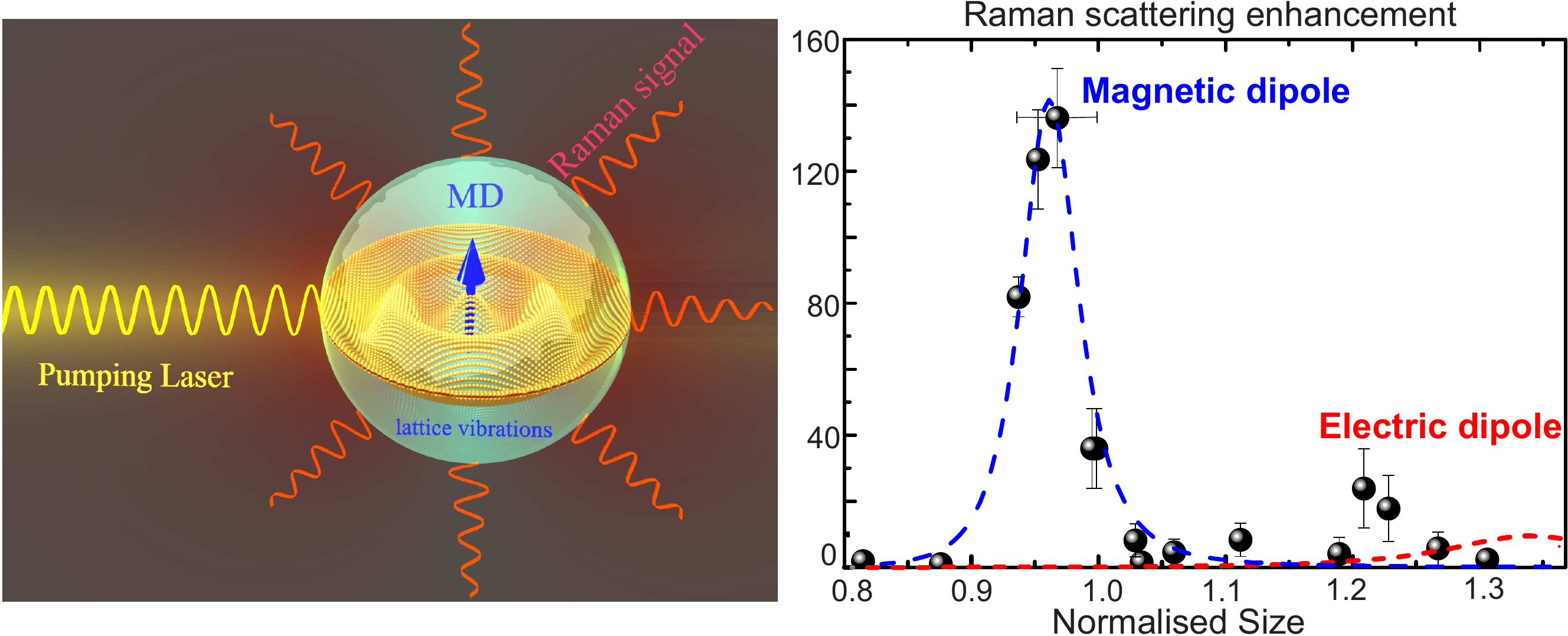}
\end{tocentry}

\begin{abstract}
Enhancement of optical response with high-index dielectric nanoparticles is attributed to the excitation of their Mie-type magnetic and electric resonances. Here we study Raman scattering from crystalline silicon nanoparticles and reveal that magnetic dipole modes have much stronger effect on the scattering than electric modes of the same order. We demonstrate experimentally a $140-$fold enhancement of Raman signal from individual silicon spherical nanoparticles at the magnetic dipole resonance. Our results confirm the importance of the optically-induced magnetic response of subwavelength dielectric nanoparticles for enhancing light-matter interactions.
\end{abstract}

Raman scattering of light is a widely used electromagnetic effect~\cite{Loudon} that offers a powerful platform for sensing~\cite{moskovits1985surface}, optical amplification~\cite{Islam}, and lasing~\cite{Pask2003}. For decades, it was well accepted that mainly metallic nanoparticles provide strong enhancement of Raman scattering via the generation of surface plasmons, being the best candidates for SERS applications~\cite{moskovits1985surface}. However, the recent advances in the study of dielectric photonic structures composed of high-index subwavelength nanoparticles pave a way towards \textit{all-dielectric resonant nanodevices} with strong field enhancement~\cite{rodriguez2014silicon, bakker2015magnetic, Caldarola2015}, large Purcell factor~\cite{Bonod2015,krasnok2015large}, and enhanced nonlinear response~\cite{shcherbakov2014enhanced, makarov2015tuning, Shcherbakov15Nano}
with low losses and low heating.

The key elements of all-dielectric nanophotonics are high-index nanoparticles~\cite{Cummer_08, Zhao09,
evlyukhin2010, kuznetsov2012, Evlyukhin:NL:2012, Miroshnichenko:NL:2012, Brener_12, krasnok2015towards} that support optically-induced Mie-type resonances in the visible without large dissipative losses inherent to metallic nanostructures. More specifically, magnetic dipole resonances have been implemented to enhance the performance of nanoantennas~\cite{Krasnok2012}, photonic topological insulators~\cite{Slobozhanyuk2015}, broadband perfect reflectors~\cite{Krishnamurthy_13}, waveguides~\cite{Savelev2014_1}, cloaking devices~\cite{cloaking2015all}, as well as dispersion control~\cite{Staude_15} and enhanced nonlinear effects~\cite{shcherbakov2014enhanced, makarov2015tuning, Shcherbakov15Nano}.  An important question is how Mie-type resonances would influence other scattering optical processes, in particular, the Raman scattering of light.

\begin{figure}[!t]
\includegraphics[width=0.7\columnwidth]{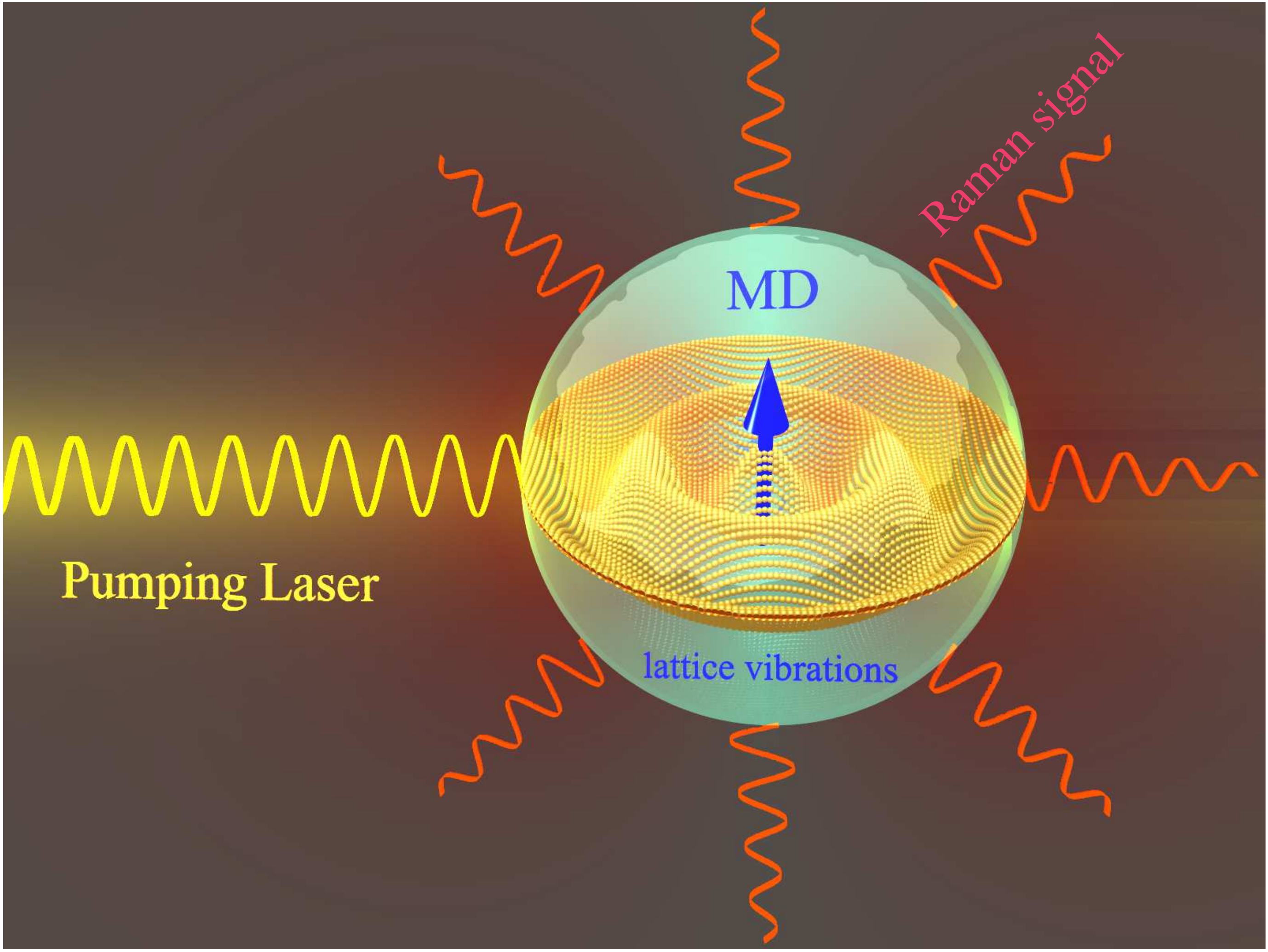}
\caption{Schematic of enhanced Raman scattering from a silicon nanoparticle
driven by an optically-induced magnetic dipole (MD) resonance.}
\label{concept}
\end{figure}

As is well known, many semiconductor materials (including crystalline silicon) demonstrate their own Raman signal contrary to noble metals~\cite{Cardona_book}. Previous studies of enhanced Raman scattering in silicon nanoparticles have been limited mainly to dense nanoparticle clusters prepared without a specific control of their shape and size~\cite{murphy1983enhanced, liu2003enhanced}, or much larger structures such as waveguides or whispering-gallery-mode resonators~\cite{spillane2002ultralow, Rong2005, Freude2010}. Here, we study Raman scattering from silicon nanoparticles in the vicinity of the Mie-type electric and magnetic resonances (see Fig.~\ref{concept}) and demonstrate a substantial enhancement of the Raman signal that depends on the size of nanoparticles. We observe experimentally almost two orders of magnitude enhancement from individual nanoparticles at the magnetic dipole resonance, and we reveal pure electromagnetic contribution into this enhancement excluding possible influence of structural effects~\cite{liu2003enhanced,meier2006raman}.

To describe the enhancement of Raman scattering and analyze the role of the electric and magnetic Mie resonances of a silicon nanoparticle, we employ the rigorous Green tensor approach. Ideologically, our theoretical approach is based on earlier related studies~\cite{Can2014, murphy1983enhanced}.

In this framework, we first determine the spatial field distribution $\mathbf{E}_{exc}(\mathbf{r})$ inside the nanoparticle at the excitation frequency created by an external source. Assuming that a spherical nanoparticle in free space is illuminated by a plane wave, we represent the normalized electric field inside the nanoparticle as a series of vector spherical
harmonics~\cite{Bohren}:
\begin{equation}
\mathbf{E}_{exc}(\mathbf{r})=\sum\nolimits_{n=1}^\infty E_n \left(c_n\mathbf{M}_{o1n}^{(1)}(\mathbf{r}) -
i{d_n}\mathbf{N}_{e1n}^{(1)}(\mathbf{r})\right) ,
\label{eq1}
\end{equation}
where $c_n$ and $d_n$ are the Mie coefficients, ${{\mathbf{M}}_{o1n}^{(1)}}$ and ${{\mathbf{N}}_{e1n}^{(1)}}$ are the orthogonal vectorial spherical harmonics, and ${E_n} = {i^n}(2n + 1)/[n(n + 1)]$. The excitation field distribution at each point inside the medium defines the Raman polarization oscillating at the Stokes frequency $\omega_s$ according to
\begin{equation}
{{\bf{P}}_s}\left( {\bf{r}} \right) = {\chi _s}\hat \alpha_j \left( {\bf{r}} \right){{\bf{E}}_{exc}}\left( {\bf{r}} \right),
\label{eq2}
\end{equation}
where $\chi_s$ is the scalar Raman susceptibility, and $\hat \alpha_j$ is the Raman polarizability tensor representing the threefold degenerate transverse optical (TO) phonon mode excitation~\cite{Ralston1970, Cardona_book}. Induced Raman polarization, in turn, produces an electromagnetic field at the observation point $\mathbf{r}_0$ given by $\mathbf{E}_s(\mathbf{r}_0) = (\omega_s^2/c^2)\int\limits_V {{{\hat G}_s}\left( {{{\mathbf{r}}_0},{\mathbf{r}}} \right){{\mathbf{P}}_s}\left( {\mathbf{r}} \right){d^3}{\mathbf{r}}}$, where ${\hat G_s}\left( {{{\mathbf{r}}_0},{\mathbf{r}}} \right)$ is the Green tensor at the Stokes frequency accounting for the Si nanoparticle and integration is carried out over the nanoparticle volume $V$. Finally, the collected signal at the point $\mathbf{r}_0$ is presented in the form:
\begin{equation} 
\begin{gathered}
  S\left( {{{\mathbf{r}}_0}} \right) = \sum\limits_j {\left\langle {{\mathbf{E}}_s^*\left( {{{\mathbf{r}}_0}} \right){{\mathbf{E}}_s}\left( {{{\mathbf{r}}_0}} \right)} \right\rangle }  = \sum\limits_j {\frac{{\omega _s^4}}
{{{c^4}}}\iint\limits_V {{d^3}{{\mathbf{r}}_1}{d^3}{{\mathbf{r}}_2}\left\langle {\hat G_s^*\left( {{{\mathbf{r}}_0},{{\mathbf{r}}_1}} \right){\mathbf{P}}_s^*\left( {{{\mathbf{r}}_1}} \right){{\hat G}_s}\left( {{{\mathbf{r}}_0},{{\mathbf{r}}_2}} \right){{\mathbf{P}}_s}\left( {{{\mathbf{r}}_2}} \right)} \right\rangle }}  \\
   = \sum\limits_j {\frac{{\omega _s^4}}
{{{c^4}}}\iint\limits_V {{d^3}{{\mathbf{r}}_1}{d^3}{{\mathbf{r}}_2}\hat G_s^*\left( {{{\mathbf{r}}_0},{{\mathbf{r}}_1}} \right){\mathbf{E}}_{exc}^*\left( {{{\mathbf{r}}_1}} \right){{\hat G}_s}\left( {{{\mathbf{r}}_0},{{\mathbf{r}}_2}} \right){{\mathbf{E}}_{exc}}\left( {{{\mathbf{r}}_2}} \right)\chi _s^2\left\langle {\hat \alpha _j^*\left( {{{\mathbf{r}}_1}} \right) \otimes {{\hat \alpha }_j}\left( {{{\mathbf{r}}_2}} \right)} \right\rangle }},  \\
\end{gathered}
\label{eq4}
\end{equation}
where summation is performed over the three degenerate TO phonon modes. Since the Raman scattering is a spontaneous process (until we enter the stimulated Raman scattering regime), the  induced polarization $\mathbf{P}_s$ is not coherent across the whole particle. Therefore, in Eq.~(3), the averaging is carried out over all possible realizations of the Raman polarization $\mathbf{P}_s$. Taking into account that the correlation length of the Raman scattering in silicon $L_c$ is of the order of tens nanometer and much less than the nanoparticle diameter, we approximate the correlation of the Raman polarizability tensors by the Dirac delta function $\left\langle {\hat \alpha_j \left( {{{\mathbf{r}}_1}} \right) \otimes \hat \alpha_j \left( {{{\mathbf{r}}_2}} \right)} \right\rangle \sim \delta \left({{\mathbf{r}_1} - {\mathbf{r}_2}} \right)$. Under this assumption, Eq.~(3) reduces to
\begin{equation}
S\left( {{{\bf{r}}_0}} \right) = \frac{{\omega _s^4}}{{{c^4}}}\sum\limits_j {\int\limits_V {{d^3}{\bf{r}}{{\left| {{{\hat G}_s}\left(
{{{\bf{r}}_0},{\bf{r}}} \right){{\hat \alpha }_j}{\chi _s}{{\bf{E}}_{exc}}\left( {\bf{r}} \right)} \right|}^2}} }
\label{eq5}
\end{equation}
\begin{figure}[!t]
\includegraphics[width=.7\columnwidth]{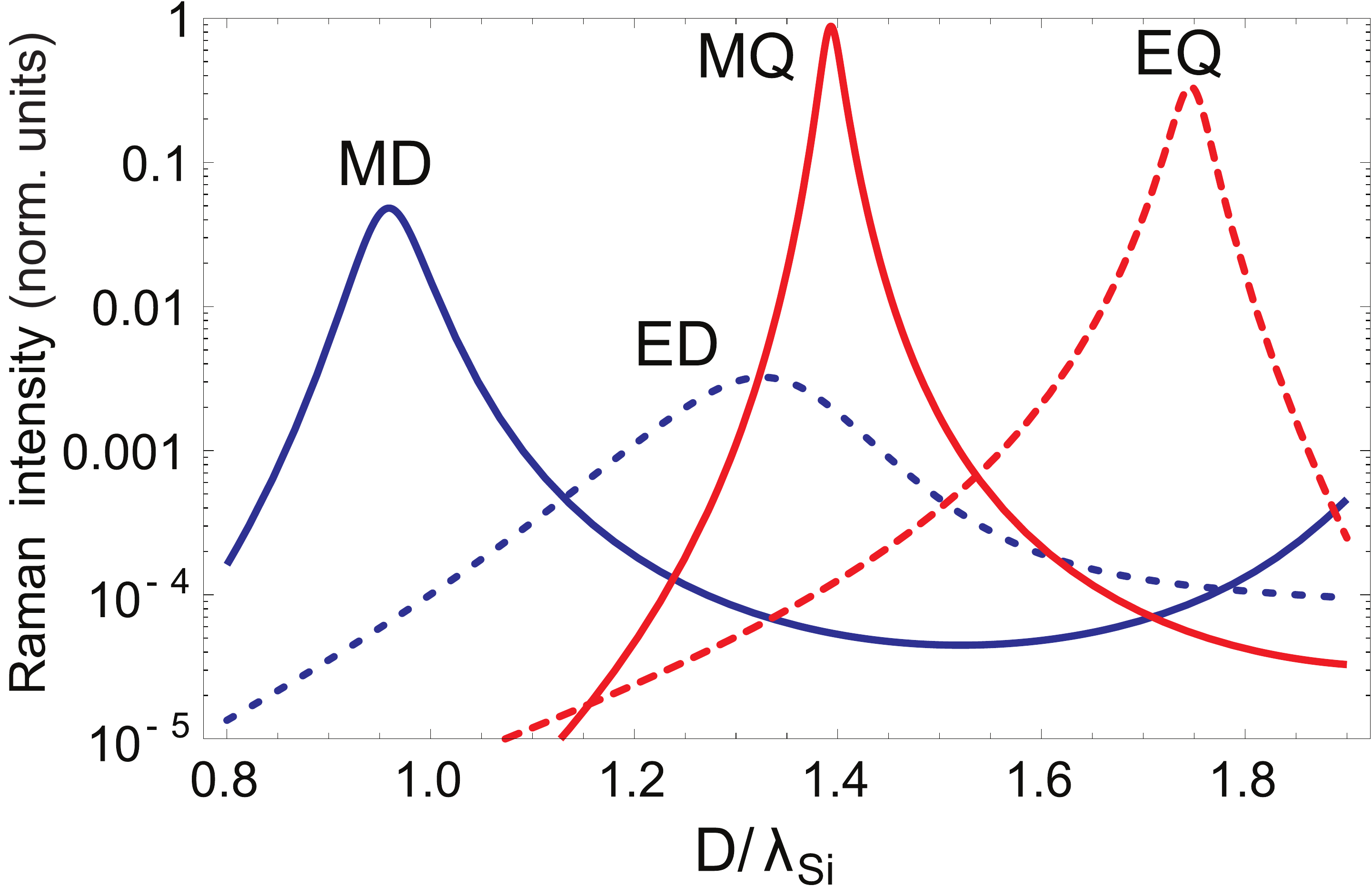}
\caption{Log plot of normalized intensity of Raman scattering as a function of dimensionless nanoparticle diameter for the magnetic dipole (MD), electric dipole (ED), magnetic quadrupole (MQ) and electric quadrupole (EQ) resonances.}
\label{fig2}
\end{figure}

Expression~(4) can be simplified with the use of the single-mode approximation. First, we notice that the electromagnetic response of an optically small Si nanoparticle at the excitation wavelength is dominated by a single magnetic or electric multipole resonance depending on the particle radius~\cite{evlyukhin2010}. Therefore, we can keep only one resonant term in Eq.~(1) and represent the electric field inside the nanopartice as ${{\mathbf{{E}}}_{exc}} \approx {E_n}{c_n}{\mathbf{M}}_{o1n}^{(1)}$, for the $n-$th magnetic resonance, and ${{\mathbf{{E}}}_{exc}}\approx -i{E_n}{d_n}{\mathbf{N}}_{e1n}^{(1)}$, for the $n-$th electric resonance, respectively. Furthermore, since the Raman shift in silicon is small compared to the linewidth $\gamma$ of Mie resonance at $\omega_0$, the main contribution to the Green tensor is provided by the same eigenmode of the system. Therefore, expanding the Green tensor in the series of eigenmodes~\cite{Ishimaru} and keeping only the resonant term, we obtain
\begin{equation}
{\hat G_s}\left( {{{\bf{r}}_0},{\bf{r}}} \right) \approx \frac{{{c^2}}}{{{N^2}}}\frac{{{\bf{u}}\left( {{{\bf{r}}_0}} \right) \otimes
{\bf{u}}^*\left( {\bf{r}} \right)}}{{{{\left( {{\omega _0} + i\gamma } \right)}^2} - \omega _s^2}},
\label{eq8}
\end{equation}
where ${{\mathbf{u}}\left( {{\mathbf r}} \right)}$ is the spatial field distribution of the eigenmode, and ${N^2} = {\int {{\mathop{\rm Re}\nolimits} \varepsilon \left( {\bf{r}} \right)\left| {{\bf{u}}\left( {\bf{r}} \right)} \right|} ^2}{d^3}{\bf{r}}$ is the normalization constant. Finally, integrating the expression (5) over the whole volume $V$ of the nanoparticle, we arrive at the following simple expression describing the Raman signal enhanced by a single Mie resonance:
\begin{equation}
S\left( {{{\mathbf{r}}_0}} \right) \approx V{\left( {\frac{{{\omega _s}}}{c}} \right)^4}{\left| {\frac{{{\chi_s s_n}}}{{{{\left( {{\omega
_0} + i\gamma } \right)}^2} - \omega _s^2}}} \right|^2},
\label{eq9}
\end{equation}
where $s_n$ stands for the Mie coefficient, either $c_n$ or $d_n$ of the corresponding mode. The above expression clearly shows that the total enhancement of Raman scattering depends on two factors: the enhancement of the excitation field inside the medium, and the Purcell enhancement of the Raman dipoles radiation~\cite{Checoury2010}.

The two key parameters entering Eq.~(6) are the resonance frequency $\omega_0$ and the resonance linewidth $\gamma$. The resonance frequency can be easily estimated numerically, while for the estimation of the resonance linewidth one can employ analytical expressions from Ref.~\cite{Lai1991}. Substituting these values into Eq.~(6), we obtain the desired spectrum of Raman scattering enhanced by the resonances of a silicon sphere. This spectrum (normalized by the particle volume $V=4\pi R^3/3$) is plotted in Fig.~\ref{fig2} as a  function of dimensionless nanoparticle diameter $D/\lambda_{\rm Si}$ with $\lambda_{\rm Si}=163$~nm being the excitation wavelength inside silicon for the magnetic dipole (MD), electric dipole (ED), magnetic quadrupole (MQ) and electric quadrupole (EQ) resonances assuming a constant excitation wavelength of 633~nm, used below in experiments.

Derived single-mode expression (6) allows us to clearly separate contributions of each Mie resonance of the nanoparticle into the total Raman scattering enhancement. As follows from Fig.~(\ref{fig2}), the strongest enhancement is associated with the MQ resonance due to its high $Q-$factor. Notably, the predicted Raman scattering enhancement at the MD resonance, which occurs for the smallest particles, is more than an order of magnitude larger than that for the ED resonance.

Theoretically predicted enhancement of Raman scattering at different Mie-resonances was directly compared with our experiments with individual crystalline silicon (c-Si) nanospheres lying on a fused silica substrate (details of the fabrication are in \textit{Methods} below). In order to determine the resonant properties of nanoparticles, we measure their scattering spectra in the dark-field scheme (Fig.~\ref{Fig3}a, see \textit{Methods} below).

\begin{figure}[!t]
\includegraphics[width=0.65\textwidth]{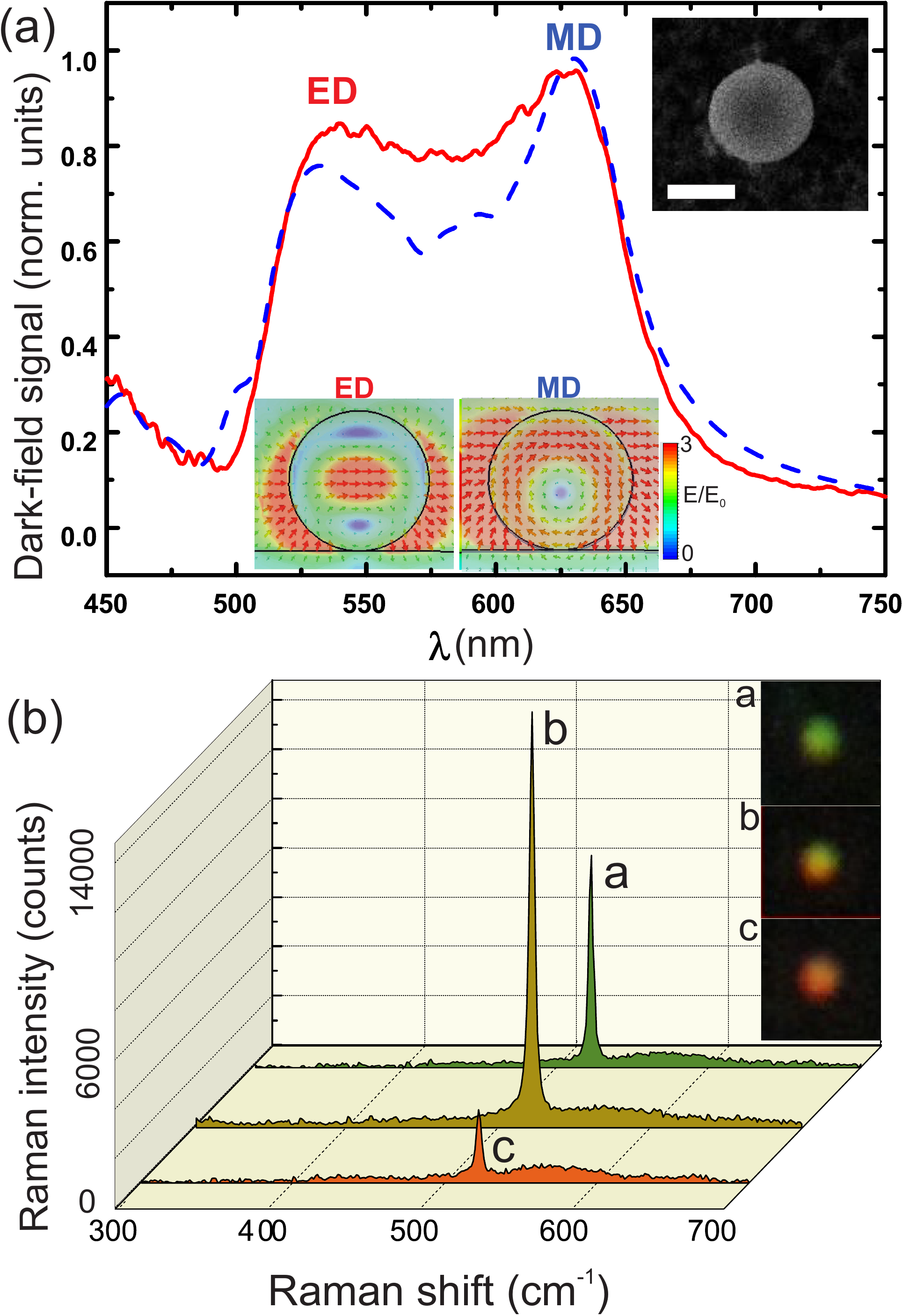}
\caption{(\textbf{a}) Experimental (solid) and theoretical (dashed) scattering spectra for s-polarized incident light. Bottom inset: the electric field distribution at different wavelengths, corresponding to electric dipole (ED) and magnetic dipole (MD) resonances. Upper inset: SEM image of typical ablative c-Si nanoparticle (scale bar represents 100~nm). (\textbf{b}) Raman spectra for different nanoparticles at the excitation wavelength of 633 nm and the corresponding dark-field optical images of the nanoparticles: (a) D=153 nm, (b) D=158 nm, (c) D=173 nm.}\label{Fig3}
\end{figure}

In order to confirm the excitation of ED and MD resonances, we simulate numerically the scattering spectra by the method of discrete-dipole approximation (DDA) and analyze near fields by the full-wave modeling in CST Microwave Studio (for details of these calculations see \textit{Methods}). The results of numerical modeling are shown in Fig.~\ref{Fig3}a, and they exhibit a good agreement with experimental results, revealing the mode structure at each spectral maximum. We use optical properties for c-Si from Ref.~\cite{VuyeSi}, giving the best fitting of our experimental scattering spectra. Minor differences of the results in the region of 550--600~nm may be attributed to the presence of a SiO$_2$ substrate.

Measurements of the Raman spectra from all fabricated nanoparticles with diameters in the range of 100-200~nm reveal a narrow peak at 521.5~cm$^{-1}$, which corresponds to TO phonons of silicon (Fig.~\ref{Fig3}b). A reference Raman signal from bulk c-Si wafer and the literature data for the Raman peak of a pure crystal exhibit almost the same phonon frequency of 520~cm$^{-1}$ with almost similar fullwidth (4--5~cm$^{-1}$)~\cite{campbell1986effects}. It is worth noting that the silica substrate does not affect Raman peak of the silicon nanoparticles, whereas confocal scheme for Raman scattering measurements allows to study response from a single nanoparticle (see Supplementary material).

\begin{figure}[!t]
\includegraphics[width=0.8\textwidth]{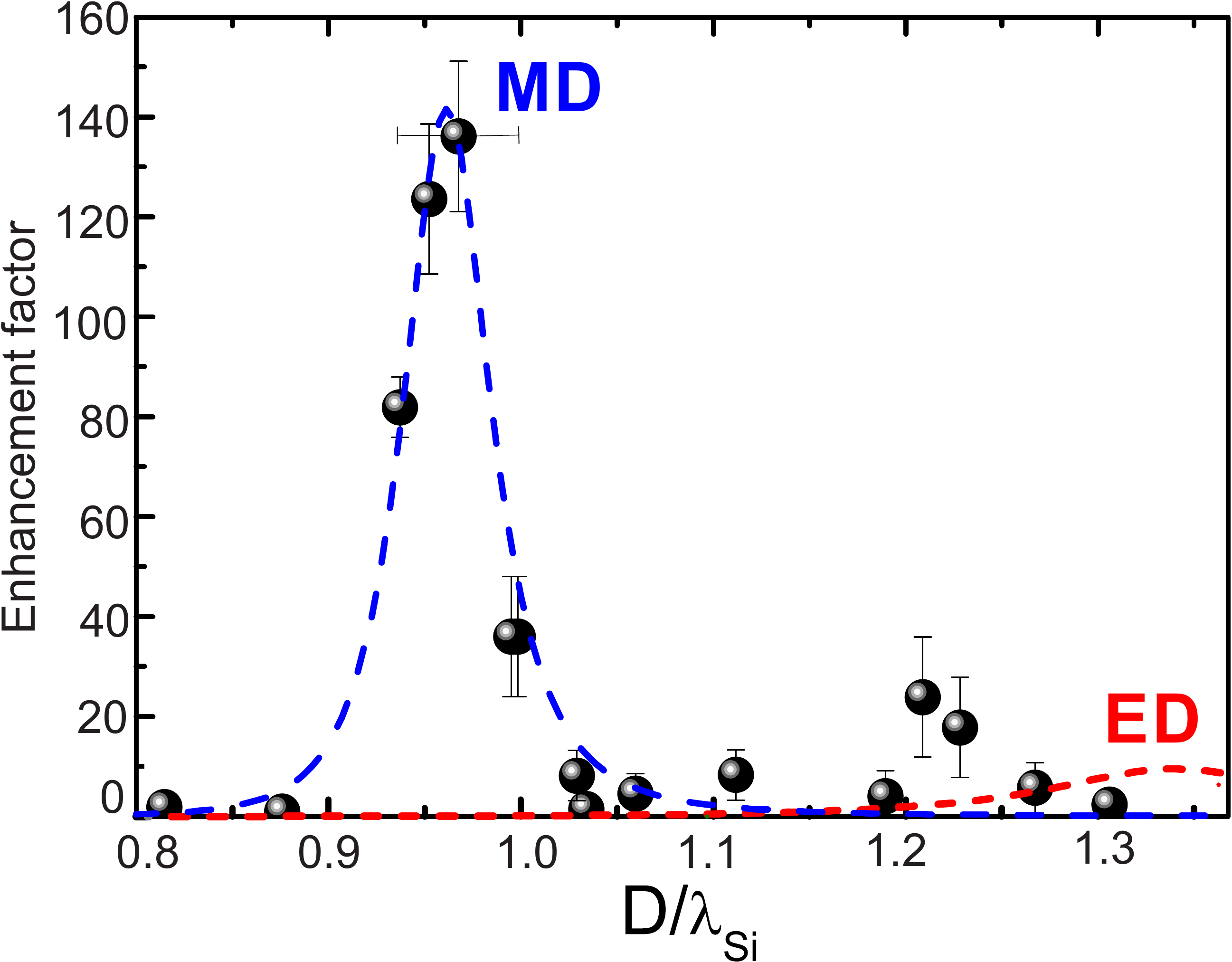}
\caption{Theoretical (dashed curves) and experimental (black dots) dependencies of the enhancement factor for Raman scattering from spherical silicon nanoparticles on their diameter D normalized to the excitation wavelength in silicon. Theoretical dependence consists of two contributions from magnetic dipole (blue dashed curve) and electric dipole (red dashed curve).}\label{Fig4}
\end{figure}

The measured Raman scattering signal from individual nanoparticles exhibits extremely strong dependence on their size and color in the dark-field images (Fig.~\ref{Fig3}b). Such a dependence for the excitation light at the wavelength $\lambda$=633~nm shows a maximum of Raman scattering for nanoparticles with $D\approx 155$~nm, supporting MD resonance at this wavelength. The maximum value of the enhancement factor (EF) for nanoparticles with MD in comparison with nanoparticles with diameters $D\approx125$~nm and $D\approx175$~nm is about $EF\approx$140. The calculation of EF from experimental data is based on the formula: EF~=~(I/I$_{\rm norm}$)$\times$(V/V$_{\rm norm}$), where I is Raman scattering signal from a studied nanoparticle with known diameter and volume V, I$_{\rm norm}$ is Raman signal from a nanoparticle of known volume V$_{\rm norm}$ with the smallest observed signal. To make such a normalization, the nanoparticle with $D\approx$135~nm is chosen. In order to check the effect of excitation wavelength, we also measured the Raman spectra at $\lambda$=532~nm for the same nanoparticles. Here, the maximum EF values observed for nanoparticles are relatively small at $\lambda$=633~nm, i.e. for diameters around 125~nm and 175~nm.

In order to distinguish contributions from each type of Mie resonances, the generalized EF dependence of Raman scattering should be represented in terms of the dimensionless nanoparticle diameter $D/\lambda_{\rm Si}$, taking into account different refractive indices at different wavelengths (Fig.~\ref{Fig4}). Such a dependence exhibits a pronounced maximum with a peak $EF\approx$140 at D/$\lambda_{\rm Si}$ $\approx$ 1, i.e. near the magnetic dipole resonance. This value is 5-7 times larger than EF for the electric dipole. Insets in Fig.~\ref{Fig3}a provide an illustrative interpretation of this enhancement. At the MD resonance, a larger farction of electromagnetic energy is stored inside the nanoparticle, thus increasing total Raman polarization and emission.
Corresponding theoretical calculations for perfect spherical c-Si nanoparticles predict even larger difference between MD and ED ($\sim$ 10), which is not perfectly matched with our observations owing to the existence of nanoscale deviations and few-nm natural oxide layer~\cite{fu2013directional, zywietz2015electromagnetic}. Nevertheless, the Raman signal enhancement in the vicinity of MD is in excellent agreement with our model. The data shown in Fig.~\ref{Fig4} is limited to nanoparticle diameters $D/\lambda_{\rm Si}<1.3$ as our fabrication method does not allow to make larger particles without pronounced ellipticity. At the same time, even small deviation from the spherical shape leads to suppression of the MQ resonance~\cite{fu2013directional}.

In conclusion, we have shown that the magnetic dipole mode of silicon nanoparticles has much stronger effect on Raman scattering as compared with the corresponding electric mode. In particular, we have observed the $140-$fold enhancement of Raman signal from individual silicon nanospheres at the magnetic resonance due to a large mode volume of the Mie-type magnetic mode compared to electric mode. Enhancement of Raman signal at the magnetic dipole resonance in high-index dielectric nanoparticles has several advantages: (i) magnetic dipole resonance is the first Mie resonance of a sphere corresponding to an ultimately compact configuration of Raman-active cavities; (ii) magnetic dipole mode is highly robust against possible deviations of the particle shape, whereas $Q-$factor of higher modes is much more sensitive to geometry; (iii) magnetic dipole mode provides at least two orders of magnitude higher enhancement of Raman scattering as compared with the non-resonant nanoparticles. Thus, our results reveal the importance of magnetic optical response of subwavelength dielectric nanoparticles for light-matter interaction and may have far reaching implications in the design of Raman lasing, bioimaging, and drug delivery applications.

\begin{acknowledgement}
This work was supported by the Russian Science Foundation: nanoparticles fabrication and optical characterization were supported by grant 15-19-00172, Raman scattering measurements and theoretical modeling were supported by grant 15-19-30023. D.G.B. acknowledges support from the Dynasty Foundation. The authors are thankful to A.~Kuznetsov, B.~Chichkov and M. Limonov for useful discussions, A. Sitnikova for TEM measurements, A.S. Gudovskikh for a-Si:H thin layer deposition by PECVD.\\

The authors declare no competing financial interest.
\end{acknowledgement}

\section{Methods}
{\em Nanoparticles fabrication.} Fabrication of c-Si nanospheres is carried out by means of the laser-induced forward transfer technique~\cite{kuznetsov2010laser} from 100 nm thick amorphous silicon film, allowing to deposit spherical nanoparticles on any substrates. A commercial femtosecond laser system (Femtosecond Oscillator TiF--100F, Avesta Poject) is used, providing laser pulses at 800~nm central wavelength, with the maximum pulse energy of 1.8~nJ, and pulse duration of 100~fs at the repetition rate of 80~MHz. The individual laser pulses are selected by means of a Pockels cell based pulse peaker (Avesta Poject). Laser energy is varied and controlled by polarization filters and a power meter (FielfMax II, Coherent), respectively, while the pulse duration is measured by an autocorrelator (Avesta Poject).

Laser pulses of 1~nJ are tightly focused by an oil immersion microscope objective (Olympus 100$\times$) with a numerical aperture NA=1.4. According to the relation d$\approx$1.22$\lambda$/NA, the estimated diameter of the beam focal spot size is d$\approx$0.7~$\mu$m, which is close to the measured value of the beam size at 1/e-level 0.68~$\mu$m by standard method~\cite{iu1982simple} based on the dependence of the laser-damaged area on the incident laser energy. In order to print nanoparticles, a laser beam is focused on a 80-nm-thick $\alpha$-Si:H film deposited on properly cleaned substrate of fused silica by the plasma enhanced chemical vapor deposition method with SiH$_{3}$ gas. The morphology of the nanoparticles is studied by means of scanning electron microscopy (SEM).

{\em Optical characterization.} All of the optical characterization is carried out on a multifuncational setup (see \textit{Supplementary materials}, Fig.~S1), which allows for precise measurements of single nanoparticle scattering. The sample is positioned by an XYZ-stage with 100~nm precision, enough to place single nanoparticles in the very center of the excitation beam. For the dark-field scattering experiments, the nanoparticles are excited at an oblique angle of incidence (65 degrees with respect to the surface normal) by linearly polarized light from a halogen lamp (HL--2000--FHSA) through a weakly-focusing objective (Mitutoyo M Plan Apo NIR, 10x, NA=0.28). The scattered light is collected from the top by an objective (Mitutoyo M Plan APO NIR, 100x, NA=0.7), sent to a Horiba LabRam HR spectrometer and projected onto a thermoelectrically cooled charge-coupled device (CCD, Andor DU 420A--OE 325) with a 150-g/mm diffraction grating. The Raman measurements are carried out by exciting the nanoparticles by one of two laser sources: a 632.8-nm HeNe laser or a 532-nm Nd:YAG laser. The exciting radiation is focused on the nanoparticles through an objective (Mitutoyo M Plan APO VIS, 100x, NA=0.9); the Raman-scattered light is collected by the same objective, the excitation light is filtered by an edge filter and sent to the same spectrometer with a 600-g/mm diffraction grating.

{\em Modeling of scattering spectra in the dark-field scheme.} The calculated geometry in the discreet-dipole approximation method is modeled as a set of interacting dipoles of c-Si, forming a silicon sphere with a given diameter on a flat fused silica substrate. Properties of the dielectric nanoparticles in the optical frequency range and the distribution of the electric fields in its vicinity are studied numerically using CST Microwave Studio. The CST Microwave Studio is a full-wave 3D electromagnetic field solver based on finite-integral time domain (FITD) solution technique. A nonuniform mesh is used to improve an accuracy in the vicinity of the Si nanoparticles where the field concentration is significantly large and inhomogeneous.


\providecommand{\latin}[1]{#1}
\providecommand*\mcitethebibliography{\thebibliography}
\csname @ifundefined\endcsname{endmcitethebibliography}
  {\let\endmcitethebibliography\endthebibliography}{}

\end{document}